\documentclass[a4paper,11pt]{article}
\pdfoutput=1

\usepackage{jheppub} 
\usepackage[T1]{fontenc}
\usepackage{pstricks}
\usepackage{color}
\usepackage{graphicx}
\usepackage{multirow}

\title{A neutrino mass-mixing sum rule from SO(10) and neutrinoless double beta decay}

\author[a]{F.~Buccella,}
\author[a,b]{M.~Chianese,}
\author[a]{G.~Mangano,}
\author[a,b]{G.~Miele}
\author[a,b]{S.~Morisi}
\author[a,b]{and P.~Santorelli}

\affiliation[a]{INFN, Sezione di Napoli, Complesso Univ. Monte S. Angelo, I-80126 Napoli, Italy}
\affiliation[b]{Dipartimento di Fisica {\it Ettore Pancini}, Universit\`a di Napoli Federico II, Complesso Univ. Monte S. Angelo, I-80126 Napoli, Italy}

\emailAdd{buccella@na.infn.it}
\emailAdd{chianese@na.infn.it}
\emailAdd{mangano@na.infn.it}
\emailAdd{miele@na.infn.it}
\emailAdd{stefano.morisi@gmail.com}
\emailAdd{pietro.santorelli@na.infn.it}

\abstract{
Minimal SO(10) grand unified models provide phenomenological predictions for neutrino mass patterns and mixing. These are the outcome of the interplay of several features, namely:  i) the seesaw mechanism; ii) the presence of an intermediate scale where B-L gauge symmetry is broken and the right-handed neutrinos acquire a Majorana mass; iii) a symmetric Dirac neutrino mass matrix whose pattern is close to the up-type quark one. In this framework two {\it natural} characteristics emerge. Normal neutrino mass hierarchy is the only allowed, and there is an approximate relation involving both light-neutrino masses and  mixing parameters. This differs from what occurring when horizontal flavour symmetries are invoked. In this case, in fact,  neutrino mixing {\it or}  mass  relations have been separately obtained in literature. In this paper we discuss an example of such comprehensive mixing-mass relation in a specific realization of SO(10) and, in particular, analyse its impact on the expected neutrinoless double beta decay effective mass parameter $\langle m_{ee} \rangle$, and on the neutrino mass scale. Remarkably a lower limit for the lightest neutrino mass is obtained ($m_{\rm lightest} \gtrsim 7.5 \times 10^{-4}$~eV, at 3 $\sigma$ level).}

\begin{document}
\maketitle
\flushbottom

\section{Introduction}

Grand Unified Theories (GUT) embed the Standard Model (SM) gauge group into (semi) simple groups of higher dimension, and provide remarkable insights on issues which are left unsolved by the - yet extremely successful -  SU(3)$_c \times$SU(2)$_L\times$U(1)$_Y$ theory. Examples of these phenomenological features are the explanation of electric charge quantization, unification of the gauge couplings at some large mass scale and a prediction for the value of the Weinberg angle. Furthermore, GUTs have a smaller set of free parameters with respect to SM, and provide nice relations among fermion masses and mixing. Finally, they share the property that all matter fields, for each generation,  can be allocated in just a few of irreducible group representations (IRR): only two in case of SU(5) \cite{Georgi:1974sy} and Pati-Salam  \cite{Pati:1973uk,Pati:1974yy} groups, and a single 16-dimensional spinorial representation for SO(10) \cite{Georgi:1974my, Fritzsch:1974nn} (for a review see ref.~\cite{Slansky:1981yr}).

In this paper we focus on the  SO(10) GUTs, pointing out that, adopting minimal and reasonable assumptions that will be discussed in the following, two interesting phenomenological implications about neutrino masses and mixing emerge:
\begin{itemize}
\item only normal $\nu-$mass ordering is allowed~\cite{Albright:2004kb,Bertolini:2006pe};
\item there is a mixing dependent  $\nu-$mass sum rule that constraints the allowed region in the plane lightest neutrino mass eigenstate ($m_{\rm lightest}$) {\it vs} effective mass parameter ($\langle m_{ee} \rangle $). This eventually affects the neutrinoless double beta decay rates.
\end{itemize} 
We know from neutrino oscillations experiments, that at least two of the three active neutrinos are massive.  However, the absolute neutrino mass scale is still unknown, as well as the mass ordering. In fact, both Normal Hierarchy (NH)  ($m_1 < m_2\ll m_3$) and Inverted Hierarchy (IH) ($m_3\ll m_1 < m_2$) are still allowed by present data~\cite{Capozzi:2016rtj,Esteban:2016qun,Forero:2014bxa}, where by definition the mass eigenstate $m_3$ is the one that maximally mixes with flavour eigenstates $\nu_\mu$ and $\nu_\tau$. 

Indeed, it is the whole paradigm of fermion mass pattern and mixing parameters hierarchies that remains a mystery, a deep question in  particle physics known as the flavour problem. In the last decades, many ideas have been put forward as attempts to address this problem, within GUTs or in different schemes. One interesting possibility is based on the idea of extending the SM gauge group to include a symmetry acting between the three families, known as {\it horizontal} or flavour symmetries. Such a symmetry could be abelian continuous~\cite{Altarelli:1998ns} or discrete~\cite{Grimus:2004hf}, and non-abelian continuous~\cite{King:2001uz} or discrete~\cite{Babu:2002dz,Grimus:2003kq}. Some years ago, it  became very popular to exploit non-abelian discrete symmetry after preliminary experimental indications supporting an almost maximal atmospheric neutrino mixing angle $ \theta_{23}$ and a small reactor angle $\theta_{13}$ at the same time (see for instance ref.~\cite{King:2014nza} and references therein). However, recent experimental data show a clear deviation from the maximality for the atmospheric angle, and indicate a not vanishing $\sin\theta_{13} \sim \lambda_C$ with $\lambda_C \approx 0.22 - 0.23$ denoting the sinus of the Cabibbo angle \cite{Olive:2016xmw}. In view of this, non-abelian flavour symmetries at the present seem to be quite disfavoured \cite{Feruglio:2015jfa}.

Typically, flavour symmetries lead to simple relations among neutrino parameters, known as the {\it mass} and {\it mixing} sum rules.  The presence of neutrino mass sum rules was first observed in ref.~\cite{Abud:2000tr} and then studied in ref.~\cite{Altarelli:2009kr}. A phenomenological classification was given in ref.~\cite{Barry:2010yk} while a more extensive analysis based on the possible neutrino mass mechanism can be found in ref.~\cite{Dorame:2011eb} (for a review on this issue see also ref.~\cite{Gehrlein:2016wlc}). On the other hand, mixing sum rules have been introduced in ref.s~\cite{King:2005bj,Masina:2005hf,Antusch:2005kw,Petcov:2014laa} (see ref.s~\cite{Ballett:2014dua,Ballett:2013wya,Gehrlein:2015ena,Agostini:2015dna,King:2013psa,Girardi:2014faa} for a recent discussion), and have strong implication for the connection between model building and experiments. It is worth stressing that in both cases, these sum rules may strongly impact the expected rate for neutrinoless double beta decays, since they give non trivial relations among the three neutrino masses or mixing parameters.

The mass and mixing sum rules  take respectively the general form
\begin{equation}
\kappa_1 m_1^h +\kappa_2 (m_2e^{-2 i \alpha})^h +\kappa_3 (m_3e^{-2 i \beta})^h=0 \,, \label{msr}
\end{equation}
\begin{eqnarray}
(1-\sqrt{2}\sin\theta_{23})&=&\rho_{\rm atm} +\lambda (1-\sqrt{2}\sin\theta_{13}) \cos\delta\,,\nonumber\\
\theta_{12} &=&  \rho_{\rm sol}+\theta_{13} \cos\delta \,, \label{mixsr}
\end{eqnarray}
where the neutrino masses $m_i$ are generally complex quantities, $\theta_{13},\, \theta_{23}, \, \theta_{12}$ are respectively, the reactor, atmospheric and solar angles, $\delta$ is the Dirac phase, and $\alpha$ and $\beta$ are the Majorana phases. Finally, $\kappa_{i}$, $\rho_{\rm atm}$, $\rho_{\rm sol}$, $\lambda$ and $h=-1,-1/2,1/2, 1$ are parameters that depend on the particular flavour scheme, as can be seen for instance in ref.s~\cite{King:2005bj,Masina:2005hf,Antusch:2005kw}.

Though relations like eq.s~(\ref{msr})~or~(\ref{mixsr}) are typically obtained within extensions of the SM based on flavor symmetries approaches it is worth observing that similar results can be also obtained in GUTs, where the approach is, so to say, {\it vertical} i.e. based on gauge symmetry paradigm, rather than horizontal. As an example, in ref.s~\cite{Abud:2000tr,Buccella:2010jc,Abud:2011mr} the following constraint has been obtained within a class of SO(10) models 
\begin{equation}\label{rel}
\frac{\sin^2\theta_{12}}{m_1}+\frac{\cos^2\theta_{12}}{m_2 \,e^{-2i\alpha}}+\frac{1}{m_3 \,e^{-2i\beta}} =0 \, .
\end{equation}
This is a mass-mixing sum rule, since the coefficients $k_{i}$ of eq.~(\ref{msr}) are functions of the mixing oscillation parameters.

In the present paper, we analise in detail the SO(10) neutrino mixing-mass sum rule in view of the latest results for the neutrino mixing parameters. Since the neutrino sum rules constrain the Majorana phases entering in the lepton mixing matrix, we examine the predictions of the SO(10) neutrino mass-mixing sum rule for the neutrinoless double beta decay. The main results are the presence of a lower limit for the lightest neutrino mass and that only normal ordering for the neutrino mass hierarchy is allowed.

The paper is organised as follows. In Section~2 we discuss the main assumptions behind the relation of eq.~(\ref{rel}). In Section~3 we study the phenomenological implications of the neutrino mass-mixing sum rule to the neutrinoless double beta decay. Finally, Section~4 is devoted to the conclusions.

\section{Mass-mixing sum rule from SO(10)}

The relation of eq.~(\ref{rel}) has been obtained in SO(10) models under the following assumptions~\cite{Abud:2011mr}:
\begin{itemize}
\item {\it type-I seesaw mechanism is dominant over type-II}.\\
In SO(10) there are two contributions to the light neutrino mass given by the type-I~\cite{Minkowski:1977sc,Yanagida:1979as,GellMann:1980vs,Schechter:1980gr,Mohapatra:1980yp} and type-II~\cite{Schechter:1980gr, Mohapatra:1980yp,Cheng:1980qt,Lazarides:1980nt} seesaw. While right-handed neutrino mass is generated at an intermediate scale $M_X\sim 10^{11}$~GeV, where the Pati-Salam group is broken, the type-II contribution is suppressed by the mass of the scalar electroweak triplet~\cite{Chang:1983fu,Chang:1984uy,Buccella:1986hn,Mohapatra:1992dx}. Such a mass is proportional to the vev of the {\bf 210} scalar IRR, which drives the breaking of SO(10) at the GUT scale. For this reason type-I seesaw is more natural in SO(10) with respect to type-II. This result has been also numerically checked and confirmed in ref.~\cite{Bertolini:2006pe}.
\item {\it In addition  to the {\bf 210}, the Higgs scalar sector contains the {\bf 10} and $\overline{\bf126}$ IRRs as well.}\\
This implies that the Dirac neutrino mass matrix $m_D$ is symmetric, and it is diagonalised by the unitary matrix $V_L$.
\begin{equation}\label{diag}
m_D=V_L^\dagger m_D^{\rm diag}V^*_L \,.
\end{equation}
\item {\it The Dirac neutrino mass matrix $m_D$ has approximatively the same structure of the up-type quark mass matrices $M_u$.}\\
This is rather a good approximation. In fact, due to the bottom-tau mass unification at the $M_X$ scale, the vev of the {\bf 10} must be dominant over the $\overline{\bf126}$ one. 
The fact that $m_D\approx M_u$ implies that the Dirac neutrino mass eigenvalues are strongly hierarchical like the up quark case, and the corresponding diagonalising matrix $V_L$ has a Cabibbo-like structure, where only the angle in the 1-2 plane is large and of the order of $\lambda_C$.
\begin{equation}\label{VL}
V_L \approx \left(\begin{array}{ccc}
\cos\theta^L_{12} & \sin\theta^L_{12} & 0 \\
- \sin\theta^L_{12} & \cos\theta^L_{12} & 0 \\
0 & 0 & 1
\end{array}\right)\,.
\end{equation}
\item {\it There is an upper limit on the mass of the heaviest right-handed neutrino $M_{R3}\lesssim 10^{11}$~GeV}.\\ 
This is related to the intermediate B-L symmetry breaking.
\end{itemize}

The minimal SO(10) model with one {\bf 10} and one $\overline{\bf126}$ IRRs were first discussed in ref.~\cite{Babu:1992ia}. We remark that such a model contains 13 free parameters to fit the charged fermion masses and the quark mixing parameters. On the other hand, the neutrino masses and mixing parameters are completely determined from the input parameters. A lot of efforts has been made in the last decades in order to check the viability of this minimal SO(10) model in view of a better understanding of the neutrino physics~\cite{Bertolini:2006pe,Babu:1992ia,Lavoura:1993vz,Lee:1994je,Brahmachari:1997cq,Oda:1998na,Matsuda:2000zp,Matsuda:2001bg,Fukuyama:2002ch,Fukuyama:2015kra}. In particular, as recently pointed in ref.s~\cite{Bertolini:2006pe,Fukuyama:2015kra}, it is possible to obtain a reasonable fit of fermion masses and mixing parameters with only type-I seesaw mechanism. The only residual discrepancy in these fits concerns the down quark mass, which is reproduced with a deviation from the {\it ``experimental''} value of about 2~$\sigma$ in ref.~\cite{Bertolini:2006pe} and 1~$\sigma$ in the more recent analysis of ref.~\cite{Fukuyama:2015kra}. It is worth observing that extending the scalar sector by adding extra {\bf 10} and $\overline{\bf126}$ Higgses would improve the global fit without spoiling the main results of our study, since the neutrino mass structure would remain the same while the number of free parameters would increase. For the sake of simplicity, in the following we focus on the minimal model with scalars belonging to a single {\bf 10} and $\overline{\bf126}$ only.

In this framework, from type-I seesaw mechanism we have
\begin{equation}
m_\nu=-m_D\frac{1}{M_R}m_D^T \,, 
\end{equation}
or
\begin{equation}\label{ss1}
M_R=-m_D^T\frac{1}{m_\nu}m_D \,,
\end{equation}
where $m_\nu$ and $M_R$ are the light neutrino and the right-handed mass matrices, respectively. By considering eq.~(\ref{diag}) and by diagonalising the light neutrino mass matrix through the neutrino mixing matrix $U$, the relation in eq.~(\ref{ss1}) can be also rewritten as
\begin{equation}\label{ss2}
M_R= - V_L^\dagger m_D^{\rm diag}V_L^*\left(U\frac{1}{m_\nu^{\rm diag}}U^T\right) V_L^\dagger m_D^{\rm diag}V_L^* = - V_L^\dagger m_D^{\rm diag} A_L m_D^{\rm diag}V_L^* \, ,
\end{equation}
where
\begin{equation}\label{ss2.0}
A_L = V_L^*\left(U\frac{1}{m_\nu^{\rm diag}}U^T\right) V_L^\dagger \,.
\end{equation}
From the assumption $m_D\approx M_u$ we have that $V_L$ is similar to the mixing matrix that diagonalises on the left  the up-type quark mass matrix. In first approximation it results in a rotation in the 1-2 plane with an angle of the order of Cabibbo one as provided in eq.~(\ref{VL}). Therefore, we get
\begin{equation}\label{matrixA}
m_D^{\rm diag} A_L m_D^{\rm diag} = 
\left(
\begin{array}{ccccc}
\left(A_{L}\right)_{11}m^2_{D1} && \left(A_{L}\right)_{12}m_{D1}m_{D2} && \left(A_{L}\right)_{13}m_{D1}m_{D3} \\
\left(A_{L}\right)_{21}m_{D1}m_{D2} && \left(A_{L}\right)_{22}m^2_{D2} && \left(A_{L}\right)_{23}m_{D2}m_{D3} \\
\left(A_{L}\right)_{31}m_{D1}m_{D3} && \left(A_{L}\right)_{32}m_{D2}m_{D3} && \left(A_{L}\right)_{33}m^2_{D3}
\end{array}
\right) \, ,
\end{equation}
where the quantities $m_{Di}$ with $i=1,2,3$ are the three eigenvalues of the Dirac neutrino mass matrix. It is worth observing that the matrix of eq.~(\ref{matrixA}) is strongly hierarchical due to the hierarchy of the mass matrix $m_D^{\rm diag}$, namely $m_{D1,2}\ll m_{D3}\sim \mathcal{O}(m_{\rm top})$. Hence, according to eq.~(\ref{VL}) the heaviest right-handed neutrino mass $ M_{R3}$ is simply given by
\begin{equation}\label{ss2.1}
 M_{R3}\equiv (M_R)_{33}\approx \left| \left(U\frac{1}{m_\nu^{\rm diag}}U^T\right)_{33} \right| \,m_{D3}^2 \, .
\end{equation}
This means that, in order to have $M_{R3}\lesssim 10^{11}$~GeV, a strong cancellation is required, which reads
\begin{equation}\label{ss3}
 \left| \left(U\frac{1}{m_\nu^{\rm diag}}U^T\right)_{33} \right|\lesssim 10^{-2}\,{\rm eV}^{-1} \equiv \varepsilon \, .
\end{equation}
Finally taking the standard Particle Data Group parametrisation for the lepton mixing matrix $U$~\cite{Olive:2016xmw}, we have
\begin{equation}
 \left| \frac{A^2}{m_1} + \frac{B^2}{m_2 e^{-2i\alpha}} + \frac{C^2}{m_3 e^{-2i\beta}}  \right|\lesssim \varepsilon \, ,
\label{eq:main}
\end{equation}
with
\begin{eqnarray}
A & = & \cos\theta_{12} \cos\theta_{23} \sin\theta_{13} e^{i \delta} - \sin\theta_{12}\sin\theta_{23}\, , \\
B & = & \sin\theta_{12} \cos\theta_{23} \sin\theta_{13} e^{i \delta} + \cos\theta_{12}\sin\theta_{23}\, , \\
C & = & \cos\theta_{13}\cos\theta_{23}\, ,
\end{eqnarray}
which reproduces the relation in eq.~(\ref{rel}) assuming $\varepsilon=0$ and $\sin\theta_{13}=0$, $\sin\theta_{23}=1/\sqrt{2}$. Notice that the relation in eq.~(\ref{eq:main}) is a generalization of the one reported in eq.~(\ref{rel}), and we will discuss in the following its phenomenological implications.

In general,  there are no theoretical predictions about the mass hierarchy even for a given neutrino mass mechanism like the type-I seesaw, but as we have already stated in SO(10) Grand Unified models only normal $\nu-$mass ordering is allowed \cite{Bertolini:2006pe}.  This can be easily understood.  In SO(10) with just a {\bf 10} and {$\overline{\bf 126}$} in the scalar sector, three fermion mass matrices ($M_u$, $m_D$ and $m_\nu$) can be written in terms of the remaining two ($M_d$ and $M_l$) as\footnote{Here we neglect the type-II neutrino mass contribution according to the considerations given above.} 
\begin{eqnarray}
M_u&=& f_u[(3+r) M_d +(1-r) M_l] \, ,\nonumber\\
m_D&=& f_u[3(1-r) M_d +(1+3r) M_l] \, ,\\
m_\nu&=& f_\nu m_D (M_d - M_l)^{-1}m_D \, ,\nonumber
\end{eqnarray}
where $f_u,\,f_\nu,\,r$ are free parameters that are functions of the vev of {\bf 10} and {$\overline{\bf 126}$} and of Yukawa matrices (see ref.~\cite{Bertolini:2006pe} for more details). If $M_d$ and $M_l$ are strongly hierarchical this will imply the same for $M_u$ and $m_D$. On the other hand, $M_d - M_l$ can be whatever, since $M_l$ and $M_d$ are quite similar. Yet, the resulting $m_\nu$ is also hierarchical and therefore, an inverted ordering is very unnatural.

This can be also seen in a different way, starting from the relation in eq.~(\ref{rel}). As pointed out in ref.~\cite{Buccella:2010jc} one gets 
\begin{equation}
\tan^2\theta_{12} = - \frac{m_1\left(m_2e^{-2i\alpha}+m_3e^{-2i\beta}\right)}{m_2e^{-2i\alpha}\left(m_1+m_3e^{-2i\beta}\right)}\, ,
\end{equation}
that gives in the IH-limit ($m_3\ll m_1<m_2$)  a solar mixing angle such that $|\tan^2\theta_{12}|\approx 1$, inconsistent with the experimental value $0.42\pm 0.07$ at 95\% C.L. \cite{Capozzi:2016rtj}.

\section{Results on neutrinoless double beta decay}

\begin{figure}[h!]
\begin{center}
\includegraphics[width=10.5cm]{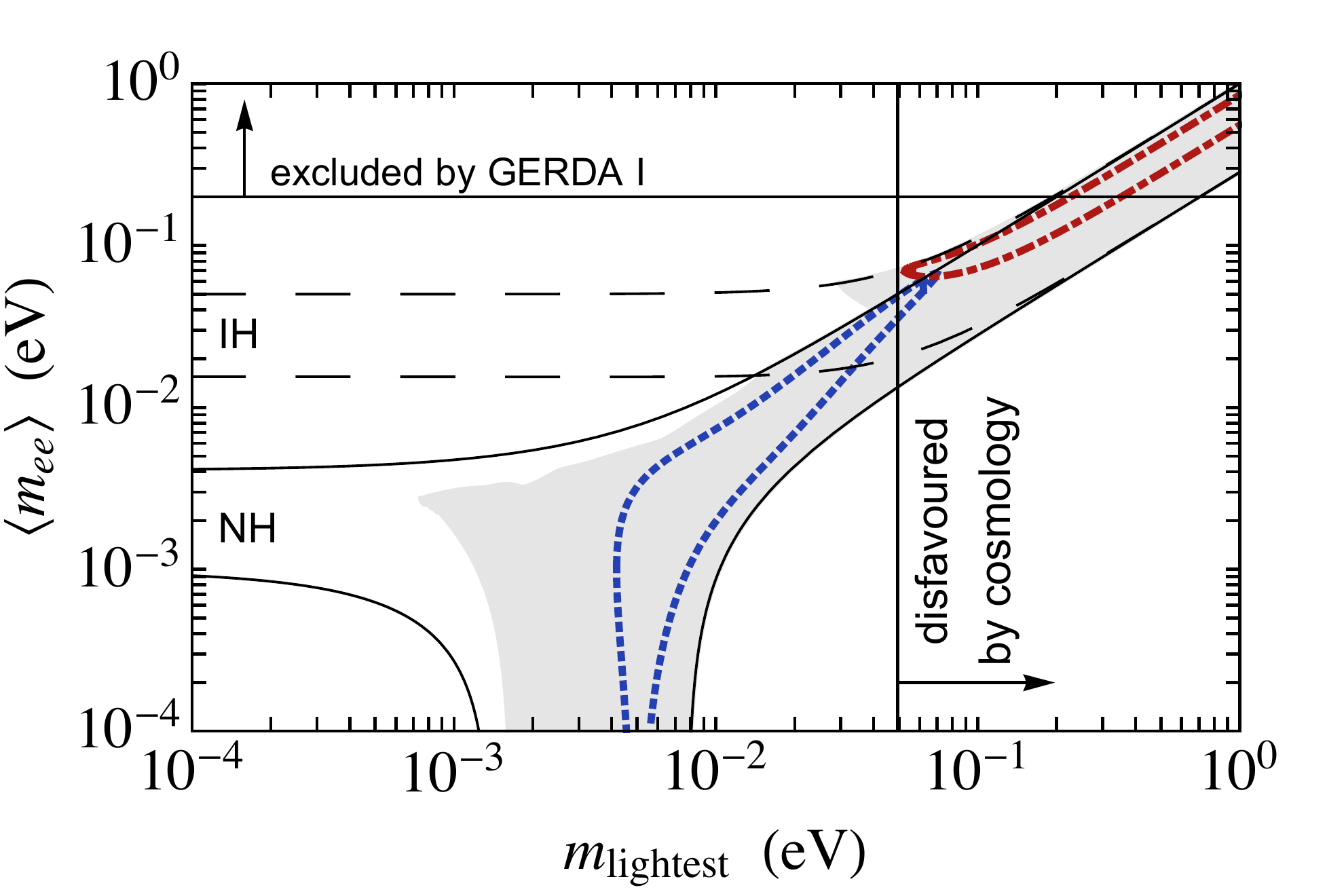}
\end{center}
\caption{\label{fig:1}The solid (dashed) lines bound the allowed region in the $m_{\rm lightest}$--$\left<m_{ee}\right>$ plane obtained by spanning the 3 $\sigma$ ranges for the neutrino mixing parameters \cite{Capozzi:2016rtj} in case of NH (IH). The dotted (dot-dashed) line is the prediction of eq.~(\ref{eq:main}) on the effective mass, once the NH (IH) best-fit values of the neutrino mixing parameters are adopted~\cite{Capozzi:2016rtj}. The shaded region represents the 3 $\sigma$ area  obtained according to the neutrino mass-mixing dependent sum rule of eq.~(\ref{eq:main}). }
\end{figure}

Remarkably, the mass relation in eq.~(\ref{eq:main}) leads to a prediction for the neutrinoless double beta decay $0\nu\beta\beta$ rates. It is well known that this decay, with a violation of the lepton number by two units, is mediated by Majorana neutrino mass terms and it would eventually demonstrate the Majorana nature of neutrinos~\cite{Schechter:1981bd}. The $0\nu\beta\beta$ decay rates are proportional to the ``effective mass''
\begin{equation}
\left<m_{ee}\right> = \left| U^2_{e1} \, m_1 + U^2_{e2} \, m_2 \, e^{2i\alpha} + U^2_{e3} \, m_3 \, e^{2i\beta} \right|\, ,
\label{eq:eff_mass}
\end{equation}
which is a function of the Majorana phases, $\alpha$ and $\beta$, and the lightest neutrino mass  $m_{\rm lightest}$, given by $m_1$ or $m_3$ in case of Normal and Inverted Hierarchies, respectively. The Dirac phase $\delta$ is included in the matrix element $ U_{e3}$. The other neutrino masses are given in terms of the measured squared mass differences, known from oscillation experiments. Due to eq.~(\ref{eq:main}) that provides a relation between the three neutrino masses and the Majorana phases one gets a bound for the allowed region in the $m_{\rm lightest}$--$\left<m_{ee}\right>$ plane, see fig.~\ref{fig:1}. In particular, the solid (dashed) lines bound the region obtained by spanning the 3 $\sigma$ ranges for the neutrino mixing parameters given in ref.~\cite{Capozzi:2016rtj} in case of NH (IH). 

In fig.~\ref{fig:1} the dotted (dot-dashed) line represents the values for $m_{\rm lightest}$ and $\left< m_{ee}\right>$ satisfying the relation~(\ref{eq:main}), once the NH (IH) best-fit values of the neutrino mixing parameters have been taken into account. Once we allow the neutrino mixing parameters to vary in their 99\% C.L. ranges one gets the shaded area reported in fig.~\ref{fig:1}. In the plot the cosmological bound $\sum m_\nu < 0.17\,{\rm eV}$, obtained by the Planck Collaboration~\cite{Ade:2015xua} (vertical line), and the constraint $\left<m_{ee}\right><0.2\,{\rm eV}$ coming from the non-observation of the neutrinoless double beta decay in the phase~1 of the GERDA experiment~\cite{Agostini:2013mzu} (horizontal line) are also shown. It is worth observing that the SO(10) relation~(\ref{eq:main}) is in agreement with an inverted hierarchy scenario only in case of a quasi-degenerate pattern. As one can appreciate from fig.~\ref{fig:1} an interesting lower limit $m_{\rm lightest} \gtrsim 7.5 \times 10^{-4}$~eV, at 3 $\sigma$ is obtained. A similar limit has been derived in left-right models with type-II seesaw~\cite{Dev:2013vxa}.

\begin{figure}[h!]
\begin{center}
\includegraphics[width=11cm]{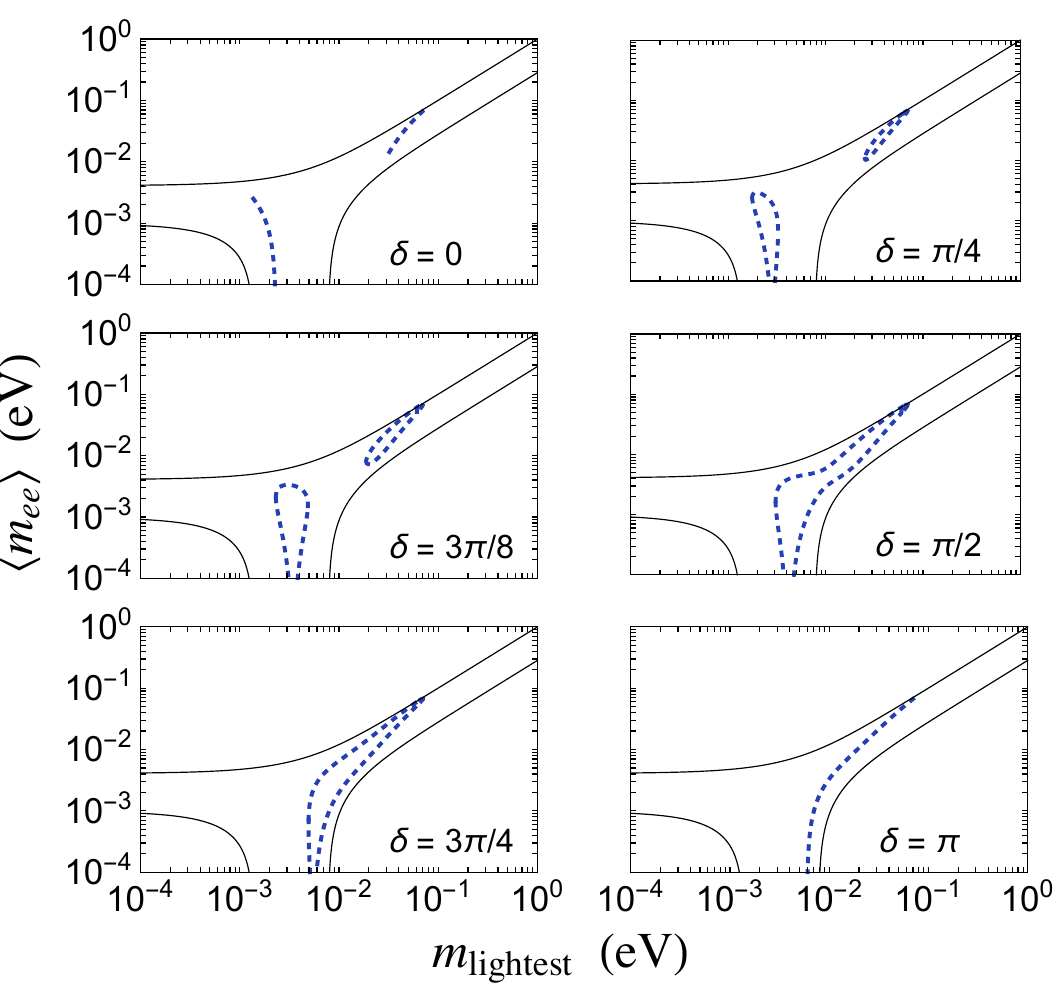}
\end{center}
\caption{\label{fig:2}The effective mass $\left<m_{ee}\right>$ from the SO(10) constraint of eq.~(\ref{eq:main}) for different values of the Dirac phases $\delta$, in the case of Normal Hierarchy. The dotted lines correspond to the allowed values in the $m_{\rm lightest}$--$\left<m_{ee}\right>$ plane when using the best-fit values of neutrino mixing parameters \cite{Capozzi:2016rtj}.}
\end{figure}

The largest uncertainty in the shaded region of fig.~\ref{fig:1} is related to the Dirac phase $\delta$, which still can range at 3 $\sigma$ in the whole range $[0,2 \pi]$ (see ref.s~\cite{Capozzi:2016rtj,Esteban:2016qun,Forero:2014bxa}). This also affects the predicted lower limit on $m_{\rm lightest}=m_1$. We show in fig.~\ref{fig:2} the best-fit lines from eq.~(\ref{eq:main}) for different choices of the parameter $\delta$ in the interval $\left[0,\pi\right]$. In particular, for small values of the Dirac phase, the relation~(\ref{eq:main}) provides two different best-fit lines that gradually merge into a single best-fit region as the parameter $\delta$ increases. When $\delta$ ranges in the interval $\left[\pi,2\pi\right]$ we get the same allowed lines shown in fig.~\ref{fig:2}, but from the bottom-right to the up-left plots. Notice that the smallest values for $m_{\rm lightest}$ are obtained for $\delta=0$.

\section{Conclusions}

In summary, we have revisited and generalized a neutrino mass-mixing relation that naturally emerges in a SO(10) GUT framework. This kind of relations has been pointed out and originally studied in the context of flavour horizontal symmetry extensions of the SM. In this paper we have rather stressed that similar relations can also arise under the spell of the gauge symmetry principle, when the SM gauge group is embedded in a larger Grand Unified Theory like SO(10), under minimal and reasonable assumptions. In this case, neutrino masses {\it and} mixing angles are involved in simple sum rules, like the one in eq.~(\ref{eq:main}), and strongly suggest a normal hierarchical pattern for neutrino masses. We have analyzed the impact of this constraint on neutrinoless double beta decay mass parameter $\left<m_{ee}\right>$, and found that a lower limit on the absolute neutrino mass scale emerges. 

So far, experiments have not been able to distinguish between the two neutrino hierarchy schemes, but there are good chances that this will be possible in the near future in several experiments, like for instance, Hyper-Kamiokande~\cite{Abe:2015zbg}, T2K~\cite{Abe:2014tzr}, ORCA~\cite{Adrian-Martinez:2016fdl}, PINGU~\cite{Aartsen:2014oha}. In this framework, a possible evidence in favour of an IH scheme will rule out the class of SO(10) models here presented. 

\section*{Acknowledgments}

The authors acknowledge support by the Istituto Nazionale di Fisica Nucleare, I.S. QFT-HEP and TAsP, and the PRIN 2012 {\it Theoretical Astroparticle Physics} of the Italian Ministero dell'Istruzione, Universit\`a e Ricerca.

\end{document}